\documentclass[epj]{svjour}

\usepackage{graphicx}
\usepackage{fancyhdr}
\def\makeheadbox{{
 }}
\usepackage{amsmath, amssymb}

\newcommand{\be}{\begin{equation}}
\newcommand{\ee}{\end{equation}}
\newcommand{\ba}{\begin{array}}
\newcommand{\ea}{\end{array}}

\newcommand{\bea}{\begin{eqnarray}}
\newcommand{\eea}{\end{eqnarray}}
\newcommand{\bma}{\begin{matrix}}
\newcommand{\ema}{\end{matrix}}
\newcommand{\bpm}{\begin{pmatrix}}
\newcommand{\epm}{\end{pmatrix}}
\newcommand{\nn}{\nonumber}
\newcommand{\half}{\frac{1}{2}}
\newcommand{\mc}{\mathcal}
\newcommand{\beq}{\stackrel{\p\mc{M}}{=}}

\newcommand{\p}{\partial}

\newcommand{\ov}{\overline}
\newcommand{\wh}{\widehat}

\newcommand{\psibar}{\ov \psi}

\newcommand{\ep}{\epsilon}

\newcommand{\la}{\lambda}
\newcommand{\da}{\delta}
\newcommand{\om}{\omega}

\newcommand{\ga}{\gamma}

\newcommand{\qrq}{\quad\Rightarrow\quad}

\newcommand{\epbar}{\ov\ep}

\def\mytitle{My title} 
\def\myauthors{My name}  
\def\mytype{My type of session}
\def\mysession{My session}

\def\mytitle{Bulk-brane supergravity}
\def\myauthors{Dmitry~V.~Belyaev}
\def\mytype{Contributed Talk}    
\def\mysession{Theoretical Models}

\begin{document}
%
%
%
%
%

\title{Bulk-brane supergravity}
\author{Dmitry~V.~Belyaev
\thanks{\emph{Email:} dmitry.belyaev@desy.de}
}
\institute{Deutsches Elektronen-Synchrotron, DESY-Theory,
Notkestrasse 85, 22603 Hamburg, Germany}
\date{}


\abstract{
We point out a limitation of the existing supergravity tensor
calculus on the $S^1/\mathbb{Z}_2$ orbifold that prevents its use
for constructing general supersymmetric bulk-plus-brane actions.
We report on the progress achieved in removing this limitation
via the development of ``supersymmetry without boundary conditions.''
} 

%
\maketitle
%

\section{Introduction}

Supergravity serves as a bridge between a more fundamental
string/M-theory on the (very) high-energy side and a variety of
``beyond the Standard Model'' extensions on the low-energy
side. If it is the only bridge, then any effective low-energy
theory of relevance to the real world should be possible to
fit into the supergravity context.

Branes and orbifolds have proven to be important (useful) in 
string/M-theory, 
and they also found very interesting realizations
at low energy. Among the latter we note the 
Randall-Sundrum scenario and  orbifold GUT models.
These models have lead to exciting research in the past decade
and are expected to be tested at the LHC. The question of fitting these
models into supergravity has also been addressed, but only partial
success has been achieved. The main purpose of this talk is to
point out some of the difficulties and indicate a possible resolution
of the related problems.

\section{Orbifold brane action}

We will concentrate on a particular setting of one-dimensional
orbifolds: $S^1/\mathbb{Z}_2$ or $\mathbb{R}/\mathbb{Z}_2$,
leading to co-dimension one fixed planes (that we, perhaps 
loosely, will call ``branes''). As our main interest is in \emph{local}
properties (local supersymmetry and boundary conditions), for most
of the discussion we need to consider only one brane.
The well-known
constructions of Horava-Witten \cite{hw} and Randall-Sundrum \cite{rs} 
belong to this class. In both cases, when one looks at the
supergravity realization of these constructions, there is a
D-dimensional bulk supergravity and some (D-1)-dimensional
brane-localized matter. 

The goal is to construct a bulk-plus-brane
action that is (locally) supersymmetric under a \emph{half} of
bulk supersymmetry (the other half being spontaneously broken by
the presence of the brane). In fact, as the bulk action is already
known (it is one of the standard D-dimensional supergravity actions),
all we need to find is the brane action. 

The brane action, in general, should include interaction between
two types of fields,
\begin{itemize}
\item
induced fields (bulk supergravity fields evaluated at the 
location of the brane), and
\item
brane-localized fields (living \emph{only} on the brane).
\end{itemize} 
These fields should combine into multiplets (representations) of the 
\emph{induced supersymmetry algebra}. As the latter may, a priori,
be \emph{different} from the standard (D-1)-dimensional supersymmetry
algebra (since the brane is \emph{embedded in the bulk} and does not
represent a closed system), 
the standard methods of constructing supersymmetric
(D-1)-dimensional actions may not be applicable to this problem.

To complicate the matters even more, the brane action, in general,
is \emph{not even separately supersymmetric}, because the supersymmetry
variation of the bulk supergravity action may produce a brane-localized
contribution which must then be canceled by the variation of the
brane action.

\section{Upstairs and downstairs pictures}

Let us consider the $\mathbb{R}/\mathbb{Z}_2$ orbifold and choose
the D-dimensional coordinates $(x,z)$ so that the brane (fixed plane) 
is at $z=0$. Orbifolding makes fields on one side of the brane be
mirror images of the fields on the other side. More precisely,
bulk fields $\Phi(x,z)$ get subdivided into two classes
of ``even'' and ``odd'' fields, 
\bea
\label{parity}
\Phi_\text{even}(x,-z) &=& +\Phi_\text{even}(x,z) \nn\\
\Phi_\text{odd}(x,-z) &=& -\Phi_\text{odd}(x,z).
\eea
Therefore, the dynamics of such a system can be completely
specified by writing a bulk-plus-\emph{boundary} action for the
fundamental domain $z\in[0,+\infty)$ which is, geometrically,
a manifold $\mc{M}$ with boundary $\p\mc{M}$. This approach
is called ``downstairs picture'' \cite{hw}.

Alternatively, one can
keep working on the total space, $z\in(-\infty,+\infty)$ (which
is a manifold \emph{without} boundary), with an additional
requirement of symmetry under the $\mathbb{Z}_2$ reflection.
This approach is called ``upstairs picture''  \cite{hw}.

The two approaches are physically equivalent, but very different
technically. In the upstairs picture, one has to deal with the
fact that odd fields are, in general, \emph{discontinuous}:
\bea
\Phi_\text{odd}(x,-0) \neq \Phi_\text{odd}(x,+0).
\eea
As we will see, \emph{on-shell}, 
the discontinuities (or ``jumps'') of the odd fields are
related to brane-localized sources, so that
\bea
\label{BC}
\Phi_\text{odd}(x,+0)=\text{brane sources}.
\eea
The brane-ortho\-gonal derivative $\p_z$ acting on the discontinuous
fields produces a brane-localized delta function, $\da(z)$.
With supergravity being a highly non-linear theory, one then
finds various products of distributions, such as
\bea
\da(z)^2, \quad \ep^2(z)\da(z), \quad \text{etc.}
\eea
in the supersymmetry transformation laws and the supersymmetry variation
of the bulk-plus-brane action. Here $\ep(z)$ is a ``sign function,''
\bea
\ep(z)=\left\{ \bma +1, z>0 \\ -1, z<0 \ema \right.
\eea
that arises as a \emph{profile function} for odd fields:
\bea
\Phi_\text{odd}(x,z)=\ep(z)\Phi_\text{odd}(x,|z|).
\eea
The products of distributions are, in general, not well-defined.
One way to make sense of them is to impose certain relations
between the distributions involved. For example, demanding
$\p_z\ep(z)=2\da(z)$ gives
\bea
\ep(z)^2\da(z)=\frac{1}{3}\da(z).
\eea
These kind of relations are indeed important for constructing
supersymmetric bulk-plus-brane actions in the upstairs picture
\cite{conrad,abn,bb1}.

On the other hand, none of this fancy mathematics is needed
in the downstairs picture, because 
on a manifold $\mc{M}$ with boundary $\p\mc{M}$ 
all fields are \emph{continuous}. 

In the downstairs picture,
it is still instructive to use the subdivision of bulk fields 
into even and odd ones. They all have now well-defined 
boundary-induced values,
\bea
\Phi_\text{even}(x,0), \quad
\Phi_\text{odd}(x,+0),
\eea
so that there is no conceptual difficulty in putting them
all in the boundary action. The equivalence with the upstairs
picture indicates only that it should be possible to find a 
boundary action that gives the same boundary condition (\ref{BC})
via the variational principle, so that, \emph{on-shell},
$\Phi_\text{odd}(x,+0)$ are fixed in terms of other fields, 
whereas $\Phi_\text{even}(x,0)$ are independent.

\section{Natural boundary conditions}

In the downstairs picture, the bulk-plus-boundary action has
the following general form,
\bea
\int_\mc{M}\mc{L}_\text{bulk}(\Phi)
+\int_{\p\mc{M}}\Big[Y(\Phi)+\mc{L}_\text{brane}(\Phi,\phi)\Big],
\eea
where $\Phi$ and $\phi$ denote the bulk and brane-localized
fields, respectively. The general (Euler-Lagrange) variation
of the action gives
\bea
\int_\mc{M}(\text{EOM})\da\Phi
+\int_{\p\mc{M}}(\text{BC})\da\Phi
+\int_{\p\mc{M}}(\text{eom})\da\phi .
\eea
Requiring this variation to vanish for \emph{arbitrary} $\da\Phi$
and $\da\phi$, gives bulk and boundary equations of motion as
well as ``natural'' boundary conditions \cite{barth}. 
For this derivation
of boundary conditions to make sense, the $Y(\Phi)$ term has to
be chosen appropriately \cite{moss,db2}. 
Its role is to bring the boundary 
variation of the bulk action to the ``$p\da q$'' form (removing
possible ``$q\da p$'' terms).
For example, the York-Gibbons-Hawking prescription,
\bea
\mc{L}_\text{bulk}(\Phi)=R \qrq Y(\Phi)=K,
\eea
brings the boundary variation to the form
\bea
\int_{\p\mc{M}}(K_{mn}-K g_{mn})\da g^{mn},
\eea
where $K_{mn}$ is the extrinsic curvature and $K$ is its trace.
The brane-localized matter adds to this variation a term
$-T_{mn}\da g^{mn}$, with $T_{mn}$ being the energy-momentum tensor,
and therefore leads to the following natural boundary conditions,
\bea
K_{mn}-K g_{mn} \beq T_{mn},
\eea
which is the downstairs picture version of the Israel matching
conditions. Here $g_{mn}$ is the induced (D-1)-dimensional
metric obtained from the bulk D-dimen\-sional metric $g_{MN}$.
In supergravity, there are more fields than just the metric.
Accordingly, the $Y(\Phi)$ term has to be extended and boundary
conditions for other fields have to be understood.

Note that this derivation of boundary conditions puts them on
the same footing as the equations of motion. On the other hand,
supersymmetry variation of a supersymmetric action must vanish
\emph{identically}, without using equations of motion. Putting
these two facts together we are led to conjecture that, if the
bulk-plus-boundary supersymmetry makes sense, it should be possible
to construct bulk-plus-boundary actions that are supersymmetric
\emph{without} using boundary conditions. Achieving this is what we
refer to as the ``\emph{supersymmetry without boundary conditions}'' 
program.

\section{Induced supergravity multiplet}

The field content of the supergravity multiplet depends on
the space-time dimension D. However, the vielbein $e_M{}^A$ and
the gravitino $\psi_M$ are always present. So, we write the
D-dimensional supergravity multiplet as
\bea
(e_M{}^A, \quad \psi_M, \quad \dots).
\eea
The supersymmetry transformations read
\bea
\label{Dsusy}
\da e_M{}^A &=& \epbar\ga^A\psi_M \nn\\
\da\psi_M &=& \p_M\ep+\wh\om_{MAB}\ga^{AB}\ep+\dots,
\eea
where $\wh\om_{MAB}$ is the supercovariant spin connection,
\bea
\wh\om_{MAB} &=& \om(e)_{MAB}+\kappa_{MAB} \nn\\
\kappa_{MAB} &=& \psibar_M\ga_A\psi_B
-\psibar_M\ga_B\psi_A+\psibar_A\ga_M\psi_B 
\eea
(all numerical coefficients are omitted). 
Splitting the D-dimensional indices into the (D-1)-dimensional ones
as $M=(m,z)$,  $A=(a,\hat z)$,
we can identify even and odd fields as follows,
\bea
\bma
\text{even:} \;& e_m{}^a \;& e_z{}^{\hat z} \;& 
\om_{mab} \;& \om_{za\hat z} \;&
\psi_{m+} \;& \psi_{z-} \;& \ep_{+} \\
\text{odd:} & e_m{}^{\hat z} & e_z{}^a & \om_{ma\hat z} & \om_{zab} &
\psi_{m-} & \psi_{z+} & \ep_{-}
\ema
\eea
where $\psi_{\pm}=\half(1\pm\ga^{\hat z})\psi$.
For the following, we will impose a gauge
\bea
e_m{}^{\hat z}=0
\eea
(using the $\la^{a\hat z}$
part of the D-dimensional Lorentz transformation)
that is very convenient \cite{db2}
in the bulk-plus-boundary setting.
Then, in particular, the extrinsic curvature is related to the
spin connection as
\bea
K_{mn}=e_n{}^a\om(e)_{ma\hat z} \, ;
\eea
$e_m{}^a$ is the induced vielbein, and $\om(e)_{mab}$ is the
corresponding torsion-free connection.

Assuming that the
unbroken half of supersymmetry is described by $\ep_{+}$,
the variation
\bea
\da e_m{}^a=\epbar_{+}\ga^a\psi_{m+}
\eea
tells us that 
\bea
\label{isg}
(e_m{}^a, \quad \psi_{m+}, \quad \dots)
\eea
should be the induced supergravity multiplet.

\section{The key point}

However, for (\ref{isg}) to be the standard (D-1)-dimensional
supergravity multiplet, the variation of $\psi_{m+}$ should
have the standard form,
\bea
\da\psi_{m+}=\p_m\ep_{+}+\wh\om^{+}_{mab}\ga^{ab}\ep_{+}+\dots,
\eea
where $\wh\om^{+}_{mab}=\om(e)_{mab}+\kappa^{+}_{mab}$ with
\bea
\kappa^{+}_{mab}=\psibar_{m+}\ga_a\psi_{b+}
-\psibar_{m+}\ga_b\psi_{a+}
+\psibar_{a+}\ga_m\psi_{b+}.
\eea
At the same time, $\wh\om_{mab}=\wh\om^{+}_{mab}+\kappa^{-}_{mab}$
with
\bea
\kappa^{-}_{mab}=\psibar_{m-}\ga_a\psi_{b-}
-\psibar_{m-}\ga_b\psi_{a-}
+\psibar_{a-}\ga_m\psi_{b-},
\eea
so that (\ref{Dsusy}) gives
\bea
\da\psi_{m+}=(\text{standard})+\kappa^{-}_{mab}\ga^{ab}\ep_{+}
+\dots
\eea
Unless the $\kappa^{-}_{mab}$ term is removed, (\ref{isg}) 
\emph{is not}
the correct (D-1)-dimensional supergravity multiplet.

This problem was resolved in Refs.~\cite{zucker,kugo} 
simply by imposing
the following boundary condition,
\bea
\label{psi0}
\psi_{m-} \beq 0,
\eea
which comes naturally with the commonly accepted ideology
that ``odd fields vanish'' at the fixed point. However,
as we will see shortly, this approach makes it impossible to
construct consistent coupling of bulk supergravity
to brane-localized matter.

\section{Do odd fields vanish?}

The standard way to argue that odd fields vanish at the
fixed point \cite{mp,bkvp} is to use both the parity condition 
(\ref{parity}) that implies
\bea
\Phi_\text{odd}(x,-0)=-\Phi_\text{odd}(x,+0),
\eea
and \emph{the assumption of continuity of fields} that gives
\bea
\Phi_\text{odd}(x,-0)=+\Phi_\text{odd}(x,+0),
\eea
from which $\Phi_\text{odd}(x,+0)=0$ does follow.
However, this argument becomes invalid in the presence of 
brane-localized sources, which, in the upstairs picture, require
odd fields to be \emph{discontinuous}. Consistency with equations
of motion leads to (on-shell) boundary conditions given in Eq.~(\ref{BC}).

\section{Boundary conditions in supergravity}

Boundary conditions must follow from (or, at least, be
consistent with) the variational principle. 
With the standard kinetic terms for $e_M{}^A$ and $\psi_M$ being
\bea
\mc{L}_\text{bulk}(\Phi)=R+\psibar_M\ga^{MNK}\p_N\psi_K+\dots,
\eea
obtaining boundary conditions from the variational principle
requires the following $Y$-term \cite{moss,db2},
\bea
Y(\Phi)=K+\psibar_{m+}\ga^{mn}\psi_{n-}+\dots
\eea
(note that it has \emph{odd} parity). This puts the boundary
part of the variation into the ``$p\da q$'' form,
\bea
\int_{\p\mc{M}} (K_{ma}-K e_{ma})\da e^{ma}
+\da\psibar_{m+}\ga^{mn}\psi_{n-} .
\eea
Brane-localized matter couples to bulk supergravity via
the induced supergravity multiplet (\ref{isg}). Therefore,
the variation of $\mc{L}_\text{brane}(\Phi,\phi)$ gives
\bea
-\int_{\p\mc{M}} T_{ma}\da e^{ma}+\da\psibar_{m+}J^m ,
\eea
where $T_{ma}$ and $J^m$ are the brane-localized 
energy-momentum tensor and the supercurrent, respectively.
This gives the following boundary conditions,
\bea
\label{TJBC}
K_{ma}-K e_{ma} \beq T_{ma}, \quad
\ga^{mn}\psi_{n-} \beq J^m ,
\eea
which is Eq.~(\ref{BC}) for the case at hand.
This makes it obvious that the boundary condition (\ref{psi0})
is allowed only when $J^m=0$. (A more strict application of
the ``odd fields vanish'' rule would require $T_{ma}=0$ as well,
which would ``kill'' even the bosonic Randall-Sundrum scenario.)

We conclude that the
orbifold supergravity tensor calculus
of Refs.~\cite{zucker,kugo} 
does not allow (consistent) construction of 
supersymmetric bulk-plus-brane actions, because the actions it
leads to are supersymmetric using the ``odd=0'' boundary conditions
which are incompatible with the ``odd=sources'' boundary conditions
following from the variational principle applied to these actions.

\section{Supersymmetry with(out) boundary conditions}

One can try to construct bulk-plus-brane actions 
that are supersymmetric using the (natural)
boundary conditions (\ref{TJBC}). 
This approach was used in Refs.~\cite{bb1,bb3} to 
supersymmetrize the
Randall-Sundrum scenario \emph{with detuned brane tensions}.
However, this road becomes very steep very soon.
The difficulty lies in the fact that as one changes the brane
action to achieve supersymmetry of the bulk-plus-brane system,
the boun\-dary conditions (\ref{TJBC}) change as well.

This is very similar to the necessity of adjusting supersymmetry
transformation laws when attempting to couple supergravity to
matter in the absence of auxiliary fields. Accordingly, one can
speculate that the proper procedure for constructing supersymmetric
bulk-plus-brane actions may include new kind of auxiliary fields,
not present in the standard (Wess-Zumino gauge-fixed) supergravity.
For example, the appearance of ``boundary compensators''
discussed in Ref.~\cite{db3} is expected.

The program of constructing bulk-plus-brane actions that are
(\emph{locally})
\emph{supersymmetric without the use of any boundary conditions}
was started in Ref.~\cite{db2}. 
There it was shown that the action for the
detuned supersymmetric Randall-Sundrum scenario of Ref.~\cite{bb1} can be
written in an alternative form so that one does not need to use
boundary conditions (\ref{TJBC}) to prove supersymmetry of the action.
However,
this statement was proven only to two-fermi order, whereas an analysis
of the supersymmetry algebra appeared to indicate that the use of
(at least) the gravitino boundary condition would be required in the 
next fermi order. 

Recent progress in this direction \cite{dbpvn}
indicates that the
program of ``supersymmetry without boundary conditions'' should
be realizable to the full extent. 
In a simpler setting of 3D supergravity,
we resolved all the problems indicated above.
We found that 
\begin{itemize}
\item
supersymmetry algebra does not impose any boun\-dary
conditions on fields;
\item
it is possible to identify
co-dimension one submultiplets, such as the induced supergravity
multiplet (\ref{isg}), without imposing any boundary conditions
on fields.
\end{itemize} 
Extending this analysis to the 5D case would improve the
orbifold supergravity tensor calculus of Refs.~\cite{zucker,kugo}  
allowing its use for 
constructing supersymmetric bulk-plus-brane actions.
The basic structure of multiplets is expected to remain unchanged 
and only be augmented by terms involving odd fields that so far
have been ``consistently'' set to zero. But even this ``minor modification''
would lead to very significant changes in the structure of the 
bulk-plus-brane actions.


\end{document}